# A Multifunctional Sub-10nm Transistor

Farshid Raissi[a], Mina Amirmazlaghani [b] and Ali Rajabi[b]

[a] *Faculty of Electrical Engineering, K. N. Toosi University of Technology, Tehran16317-1419, Iran.*
[b] *Nano Electronics Lab(NEL), Shahid Rajaee Teacher Training University, Tehran 16785-163, Iran*

ABSTRACT

Nano-electronic integrated circuit technology is exclusively based on MOSFET transistor due to its scalability down to the nanometer range. On the other hand, Bipolar Junction Transistor (BJT), which provides unmatched analog characteristics and frequency response, cannot be scaled to nanometer regime without the loss of transistor action. Here a versatile nanoscale transistor is introduced that provides identical BJT behavior and expands its capabilities. The new transistor uses CMOS fabrication technology and creates BJT's emitter, base, and collector via electric fields. By allowing carrier modulation during operation, its current gain can be changed at least by five orders of magnitude. This property introduces novel adaptive, variable gain, and programmable analog modules into existing electronic circuit design and manufacturing. A NOT gate version of this device with the critical dimension of 7 nm operates at 730 GHz, and its three-stage ring oscillator exhibits a frequency of 240 GHz. With proper gate biasing, it can also operate as a nanoscale MOSFET, easily alleviating short-channel effects.

INTRODUCTION

The incredible growth of the electronic industry in the past decades is largely due to the consistent performance of MOSFET transistors at ever-decreasing technology nodes down to sub-10nm dimensions. On the other hand, BJT is essentially left unused because its performance deteriorates as its dimensions are decreased down to nanometer scale [1-4]. BJT, however, provides analog characteristics that are unmatched by MOSFET and could enhance the performance of analog circuits if it could seamlessly be incorporated into the MOSFET fabrication process. Much work has been dedicated to developing BJT fabrication processes compatible with MOS technology even if its dimensions are not at the nanometer scale to take advantage of its valuable characteristics [5-24].

Here a versatile transistor is demonstrated that is fabricated by MOS processing technology that not only provides BJT characteristics but it does it at the nanometer scale, which results in unprecedented fast operational speed and large current gain. A unique property of this transistor is that its current amplification can be changed at least by five orders of magnitude during operation. Furthermore, with proper biasing, it can operate identical to MOSFET while easily solving the so-called short-channel effects that MOSFETs encounter at the nanometer scale.

The paper is organized as follows. First, the fact that regular BJTs cannot be downscaled below the 100nm range is examined. Then the transistor proposed here is introduced, and its operation and characteristics are provided in detail. The ability to change the current amplification factor and its ramifications are presented next. Its speed is examined via a regular NOT gate and a three-stage ring oscillator in the following section. Its MOSFET behavior is examined afterward, along with its ability to reduce the size and power consumption of analog integrated circuits.

LIMITS TO BJT DOWNSCALING

BJT is comprised of three oppositely doped semiconductor areas called emitter, base, and collector regions. Its operation is based on the injection of carriers from the forward biased emitter-base junction into the reverse-biased base-collector junction [26-27]. The so-called transistor action, or basically control and amplification of signals, is caused by the charge neutrality condition at the base. In an npn transistor, there are two depletion regions extending into the p base region from n regions to its sides. The area between these two regions is considered charge neutral. Under normal operating conditions, the np junction is forward biased, and electrons are injected from the emitter to the base. These electrons can pass through the charge-neutral region of the base reaching the pn depletion region, where they are swept by its electric field into the collector. Charge neutrality means that for each injected electron, there must be a hole coming into the base to compensate for the extra electron's charge. The ratio between the time a hole spends in the base before recombination and the time it takes for the electron to traverse the neutral region determines how many electrons are coming in for each hole. The neutral region width, therefore, plays a crucial role in creating control and also the amount of current gain [1-4], [25-26]. If there is no neutral region at the base, meaning the two depletion regions reach each other, transistor action and control are lost. This could happen at nanoscale base widths for regular doping levels.

To demonstrate this, consider Fig. 1.a and b, as an example, where the widths for the base are 500nm and 100nm for the same impurity doping levels, respectively. Here TCAD-3D simulator has been used to obtain carrier densities for impurity doping levels of $10^{18}/cm^3$, $10^{16}/cm^3$, $10^{14}/cm^3$ for three npn regions, respectively. As is observed in Fig. 1.a, there are three distinct npn regions creating the emitter, base, and collector of the transistor. There is a stark difference in carrier densities of this structure and that of Fig. 1.b, where the base width is only 100nm. There is no p region at the center. This structure won't operate as a BJT. This is the main reason why BJTs cannot be downscaled. There are possibilities to decrease the base further either by doping the base much more or using heterojunction BJTs [5-9], [11-24], but these attempts are either not compatible with standard integrated circuit design and manufacturing or introduce further operational problems.

The reason for the base being completely depleted in Fig. 1.b is that electric potential needs to be dropped in order to change a region from n to p, and such potential drop is not provided by the size of the device and its impurity levels. The voltage drop across a junction is obtained by integrating impurity levels over a distance twice. For this example, with the doping levels used, a voltage drop of 0.84V is necessary to change the material from p to n or vise versa. While 500nm is enough for these impurity levels, 100nm is too small a distance to create enough voltage and cause the change.

PROPOSED TRANSISTOR STRUCTURE

The proposed device is a natural progression of the Field Effect Diode (FED) [27-31]. FED is a diode whose n and p doping is provided by electric fields of two separated gates. Its structure is similar to a MOSFET but has two gates over the channel, and the doping of source and drain are opposite to each other. Nanometer size FEDs have been proposed for various circuit applications [32-41]. It has been proven that FED performs identically to a regular pn junction diode exhibiting carrier injection, the exponential IV, and the characteristic reverse bias behavior.

The natural step forward is to examine a BJT version where three gates create emitter, base, and collector regions inside the channel or a structure hereafter called field effect BJT (FEBJT). The schematic representation of FEBJT is provided in Fig.2. As is observed, there are three gates over the channel. To create an npn FEBJT, the two side-gates are biased positive and the center gate negative with respect to the semiconductor. Connections to the channel are provided along the widths of the two side-gates and along the length of the center gate. The dimensions given are those used in simulations. The length of the center gate, or the *base*, is 7nm. The separation between the gates is 7nm as well. The two side-gates are 10nm in length. The length of the two side-gates is not critical because the speed and gain of the device are mostly dependent on the length of the base gate.

Again TCAD-3D was used to obtain the band diagram along with the channel and carrier densities, and the results are provided in Fig. 3 and 4. The parameters used and other relevant information are provided in table 1. As both figures indicate, three clear areas of n, p, and n are created inside the channel by the gate voltages. This is to be compared to Fig. 1.b, where a base of 100nm was not enough to provide the necessary impurity charge to change the n region into p. Here over an area of less than 10nm, the center p region has been created. The reason for this difference is their doping mechanisms. One uses impurities to create electrons and holes, and the other uses electric fields set from outside the semiconductor. Impurities cannot create electric potential necessary for changing an n region to p if they are not given enough distance, while FEBJT uses gate voltages to generate enough electric field and voltage drop inside the semiconductor without a restriction on the size or distances between the gates.

Based on the equality of FED behavior with regular pn junction diode, one can theoretically extend its underlying mechanisms to FEBJT and conclude that it must operate in an identical fashion to BJT.

Following simulations corroborate this fact. The IV curves in the forward-active mode are presented in Fig. 5. Forward-active mode refers to the case when the emitter-base junction is forward, and the base-collector junction is reverse biased. IV characteristics present collector currents versus collector to emitter voltage for different base currents. From these graphs existence of carrier injection from emitter to collector, control of collector current by the base current, and current amplification factor are deduced. As is observed, the very fact that current increases at low voltages and then saturates is an indication of carriers being injected into the reverse-biased base-emitter junction. On the other hand, the increase in saturation current with base current is an indication of controllability of collector current. The ratio of saturated collector current to the base current is the current amplification factor, which in this particular case is 100. Therefore FEBJT is a multi-terminal transistor capable of providing current amplification similar to regular BJT transistors. Its current gain and speed are examined and discussed later.

Besides the IV characteristics of Fig.5, which provides the evidence for transistor action, a separate graph called Gummel plot is normally provided for BJT structures. This graph provides information about the physical processes occurring inside the device. Gummel plot is the benchmark approach to assessing BJT behavior. It depicts collector and base current versus emitter-base voltage on a semi-log graph. The current amplification factor is obtained from the ratio of the collector to the base current, and the slopes of the two current graphs provide clues about the underlying physical processes. Gummel plot has been obtained for gate voltages that provide a typical current amplification factor of 100 and is given in Fig.6.

There are three regions with different slopes associated with each curve. The rather flat region at low voltages is due to base current recombination and is expected to happen in any regular forward bias diode and, therefore, each BJT.

The linear region between 0.3V and 0.8V indicates an exponential relationship between collector current and emitter-base voltage as expected for a forward bias diode. The slope at this region is *qV/nKT*, where q is an electron's electric charge, V is voltage, K is Boltzmann's constant, T denotes temperature, and n is the ideality factor. For a diode with an ideality factor equal to 1, this slope equals 60mV/decade. FEBJT graph shows 55mV/decade, indicating an ideality factor of roughly equal to 1.09. Ideality factor larger than one indicates that generation and recombination currents are also contributing to the overall current. This is acceptable because the simulation program does take into account such processes.

The roll-off at high voltages is particularly informative. It happens in regular BJTs because of a phenomenon known as a high-level injection. Among its results and ramifications is that charge neutrality is at play in the transistor. In a regular BJT, as more and more electrons are injected into the base, more and more holes come in to keep the charge neutral. The extra holes increase the current diffusing from base to emitter. This decreases the portion of emitter current that reaches the collector, and as a result, the increase in voltage does not increase the collector current as before.

In FEBJT, however, the notion of charge neutrality must be changed because the amount of charge inside the semiconductor is not set just by impurities, but charges accumulated at the gates must be taken into account as well. The gate charges, which are caused by voltages applied to them, force a certain amount of charge to exist in the base, and this value must remain constant when carriers are injected from emitter to base during FEBJT operation. This "charge constancy" operates as charge neutrality in regular BJTs. This is corroborated by the shape of the Gummel plot and is deduced from the fact that the base current is, in fact controlling the collector current and provides amplification as well.

Two very interesting and consequential features of the FEBJT are apparent right away. First, its current gain can be very large due to its unprecedented short base. Second, the carrier density of the emitter, base, and collector can be changed independently during operation as desired. These two properties are examined in more detail next as it provides a variable gain transistor covering an unprecedented dynamic range.

VARIABLE GAIN TRANSISTOR

The concept of variable gain amplifier circuits is well understood for its various practical applications, including audio level compression, synthesizers, and amplitude modulation. But the concept of a variable gain transistor is new because regular MOSFETs or BJTs have their gains built into their design and fabrication parameters that cannot be changed once fabricated. But FEBJT's current gain can be varied readily with a large dynamic range by changing the emitter and base carrier densities via their gates. Current amplification factor is defined as the ratio of the collector current to base current, referred to as the current gain written as $\beta=I_C/I_B$.

IV characteristics and Gummel plot indicate that the mechanism for current amplification for FEBJT is the same as regular BJTs. This means that the ratio of hole recombination time at the base to the transit time of electrons across the base gives the "maximum" current gain. There are other factors affecting the current gain, and for the case when the widths of the emitter, the base, and the collector are comparable with the diffusion length of electrons and holes, the following expression is used for analysis [42-43].

$$\beta = \frac{D_B W_E N_E}{D_E W_B N_B} \tag{1}$$

Where, $D_B$ and $D_E$ are the minority carrier diffusion coefficients in the base and emitter, $W_B$ and $W_E$ are the effective widths of the base and emitter, $N_B$ and $N_E$ are carrier density in the base and emitter, respectively. Effective widths refer to areas outside the depletion regions where the electric field is considered zero. Ignoring effective widths, for now, the current gain is linearly dependent on the ratio of the emitter to base carrier densities. In Fig.3, carrier densities inside the channel for three gate voltages of 1.2V, 0.8V, and -0.5V were provided, which correspond to carrier densities of $3\times10^{18}/cm^3$, $5\times10^{17}/cm^3$, and $4\times10^{12}/cm^3$, respectively. Assuming the emitter gate is biased with 1.2V and the base gate is changed from -1.2V to -0.8V and then -0.5V, the current amplification factor should change by two and then six orders of magnitude considering the carrier density ratios. However, effective widths play a role, and their determination is subjective in our device, where the edges of depletion regions are not well defined. $\beta$ values obtained by simulation are provided in the table.2 when the emitter gate voltage is kept at 1.2V. A change of 5 orders of magnitude for $\beta$ is obtained, which shows the analytical expression overestimates the values.

The unprecedented $\beta$ values of table 2 are brought about by the nanoscale size of the base. The ability to change the gain by five orders of magnitude makes this device attractive for practical applications of variable gain amplifiers that were mentioned before. On the other hand, the small size and ease of operation open up new application fields for variable gain FEBJT. These areas include novel field-programmable amplifying modules, adaptive control systems, neural control systems, non-uniformity compensation, environmental adjustment circuits, analog computing, wide-range analog to digital, and digital to analog circuits.

SPEED AND FREQUENCY RESPONSE

As a measure of FEBJT speed and frequency response, an inverter amplifier or digital NOT gate is examined. Fig.7 demonstrates the circuit in which the input is applied to the base of both transistors, and output is obtained from their common collector.
Here, the emitter and collector gate voltages are set to create an npn BJT current sink and a pnp BJT current source. Fig.8 gives the output as a square wave is applied to the input. The NOT gate propagation delay ($\tau_p$) is defined as the average of the low-to-high ($\tau_{pLH}$) and the high-to-low ($\tau_{pHL}$) propagation delays, which are defined as the times it takes for the output voltage to reach the middle between the high and low voltage levels. In our case, this is the zero voltage. The equations for propagation delay and frequency response and their values become [44-45]:

$$\tau_p = \frac{\tau_{pHL}+\tau_{pLH}}{2} = 0.68 \text{ psec} \qquad (2)$$

$$F_{max} = \frac{1}{(\tau_{pHL}+\tau_{pLH})} = 730 \text{GHz} \qquad (3)$$

Power delay product and dynamic power using the input capacitor of a subsequent inverter stage as the output are calculated as [44-45]:

$$PDP = C_{load} \times V_{dd}^2 = 0.153 \text{ fF} \times (2)^2 = 61.2 \times 10^{17} \text{ J} \qquad (4)$$

$$P_D = \text{dynamic power} = P_D = C_{load} \times V_{dd}^2 \times F_{max} = 446 \text{ } \mu W \qquad (5)$$

The reason for the very high-frequency response is the mechanism with which BJTs turn on and off. Turning on or off is basically related to the build-up of excess carriers in the base and their subsequent removal. When the device turns off, excess carriers are removed by two mechanisms of diffusion and recombination. For a base width of 7nm, it is expected for diffusion to dominate and dictate the frequency response. The momentum of carriers determines their diffusion rate, which is approximately the Fermi momentum. In a MOSFET, carriers are swept across the channel by their drift velocity that is normally less than Fermi speed.

To actually observe the ac response of FEBJT, a three-stage ring oscillator made of the same NOT gate was tested. Fig.9 shows the circuit, and Fig.10 demonstrates the output. At the initial stage of simulation, bias voltages to the side-gates were applied as a ramp to start oscillation. The frequency of oscillation is about 245 GHz in the figure.

$$F_{osc} = \frac{1}{2n\tau_p} = 1/(2 \times 3 \times 0.68 \times 10^{-12}) = 245 \text{ GHz} \qquad (6)$$

MOSFET CONFIGURATION

FEBJT is basically a generic MOSFET with the exception of three gates over its channel. If all the gates are biased with the same polarity, MOSFET-like behavior is obtained. Such a MOSFET structure, as given in Fig.11 does not include sidewalls around the gate and halo impurity implants that are normally provided to remedy short-channel effects. The fact is that this structure does not need any of those structural features to combat short-channel effects. The two side-gates, when independently biased can help reduce drain-induced barrier or hot electron effects. They also eliminate overlap between the center gate and drain and source, while their own overlap does not hinder the frequency response of the device. These eliminate several steps in manufacturing. To demonstrate the ability of the side-gates in eliminating the short channel effect, Fig.12 is provided. In this figure, the device has been biased like a MOSFET but with different side-gate voltages. At low side-gate voltages, short channel effects are apparent. Short-channeled effects are basically the hot electron phenomenon and drain-induced barrier lowering. The former causes an increase in drain current at higher drain voltages, and the letter causes an overall increase in current with drain voltage. The gate loses its ability to control the current, and saturation of current is not achieved. As seen in the figure, at high enough side-gate voltages, the normal MOSFET IV curve with its characteristics saturation region is exhibited, and both short-channel effects are eliminated.

OUTPUT STAGE CURRENT BOOSTER

Fabrication compatibility of this device with regular CMOS process along with its huge current amplification can make it ideal as the output stage of CMOS amplifying circuitry. The output stages of analog CMOS circuits have to charge and de-charge output capacitors, and the larger the output current, the faster the so-called slew rate and frequency of operation. Normally large W/L ratios are used to provide large currents at the output. Consequently, the prior stages have to provide large currents to charge the output stage fast enough, which means they also need large W/L ratios. FEBJT circuits such as that of Fig.9 can be placed at the output stage of any CMOS analog circuit that needs high currents, and the current of the prior stage can be fed to its input, which can amplify it by several orders of magnitude. This means that CMOS circuits can be designed with orders of magnitude less current or W/L ratios. Only FEBJT at the output stage needs to be of the same size as the transistors of the original CMOS circuit.

CONCLUSIONS

Using electric fields of three gates to create emitter, base, and collector regions of a BJT inside the channel of regular MOSFET results in a transistor that operates at hundreds of GHz frequency range when its base width is less than 10nm. It can operate identical to BJT transistors with the added possibility of changing emitter, collector, and base carrier densities independently. This provides the possibility of changing transistor gain by several orders of magnitude, which presents an unprecedented adaptive device useful in several applications. This device is manufactured with the same fabrication processes used by regular MOS technology and can be incorporated into analog CMOS circuits to improve their performance or decrease their size and power consumption. It can also be used similar to regular MOS transistors, with the side-gates acting to remedy short channel effects at the nanometer scale. All and all, this versatile device brings together the advantages of CMOS and BJT transistor characteristics with added capabilities for novel applications in adaptive devices and circuits.

References:

[1] Brinkman, W. F., Haggan, D. E., Troutman, W. W. A history of the invention of the transistor and where it will lead us. IEEE J. of Solid-State Circ., 32, 1858-1865 (1997)
[2] Kilby, J. Invention of the Integrated Circuit. IEEE Trans. on Elec. Dev. 7, 648–654 (1976)
[3] Balestra, F. Nanoscale CMOS. New Jersey: John wiley & sons (2017)
[4] Streetman, B. G., Banerjee, S. Solid State Electronic Devices. New Jersey: Prentice Hall (2000)
[5] Washio, K. SiGe HBT and BiCMOS technologies for optical transmission and wireless communication systems. IEEE Trans. Electron Dev., 50, 656-668, (2003)
[6] Athanasiou. S., Athanasiou, S., Legrand, C. A., Cristoloyeanu. S., Galy, P. Novel Ultrathin FD-SOI BiMOS Device With Reconfigurable Operation," IEEE Trans. Electron Dev. 64. 916-922 (2017)
[7] Asada, S., Satoshi, T., Kimoto, T., Suda, J. Design Criterion for SiC BJTs to Avoid ON-Characteristics Degradation Due to Base Spreading Resistance. IEEE Trans. Electron Dev. 64. 2086-2091 (2017)
[8] Jung., Y. et al. A novel BJT structure implemented using CMOS processes for high-performance analog circuit applications. IEEE Trans. Semicond. Manuf. 25. 549-554. (2012)
[9] Loan, S. A., Loan, S. A., Quteshi, A., Kumarlayer. S. S. A novel high breakdown voltage lateral bipolar transistor on SOI with multizone doping and multistep oxide. Semicond. Sci. Technol. 24 (2009)

[10] Chakraborty, S., Malik, A., Sarkar, C. K., Rao, V. R. CMOS Devices and Circuits for Ultralow Power Analog/Mixed-Signal Applications. IEEE Trans. Elctron Dev. 54, 241 - 248 (2007)

[11] Liu, Liwei, Ningsheng Xu, Yu Zhang, Peng Zhao, Huanjun Chen, and Shaozhi Deng. "Van der Waals Bipolar Junction Transistor Using Vertically Stacked Two-Dimensional Atomic Crystals." Advanced Functional Materials 29, no. 17 (2019): 1807893.

[12] Jha, Aashu, Thomas Ferreira De Lima, Hooman Saeidi, Simon Bilodeau, Alexander N. Tait, Chaoran Huang, Siamak Abbaslou, Bhavin Shastri, and Paul R. Prucnal. "Lateral bipolar junction transistor on a silicon photonics platform." Optics Express 28, no. 8 (2020): 11692-11704.

[13] Su, Bao-Wang, Bin-Wei Yao, Xi-Lin Zhang, Kai-Xuan Huang, De-Kang Li, Hao-Wei Guo, Xiao-Kuan Li, Xu-Dong Chen, Zhi-Bo Liu, and Jian-Guo Tian. "A gate-tunable symmetric bipolar junction transistor fabricated via femtosecond laser processing." Nanoscale Advances 2, no. 4 (2020): 1733-1740.

[14] Bédécarrats, Thomas, Philippe Galy, Claire Fenouillet-Béranger, and Sorin Cristoloveanu. "Investigation of built-in bipolar junction transistor in FD-SOI BIMOS." Solid-State Electronics 159 (2019): 177-183.

[15] Hashemi, Pouya, Bahman Hekmatshoartabari, Alexander Reznicek, Karthik Balakrishnan, and Jeng-bang Yau. "Lateral bipolar junction transistor with dual base region." U.S. Patent Application 15/828,152, filed May 30, 2019.

[16] Hung, Ching-Wen, Meng-Chi Chiang, and Yen-Chih Lin. "Bipolar junction transistor." U.S. Patent 10,529,837, issued January 7, 2020.

[17] Zhang, Yourun, Hang Chen, Maojiu Luo, Juntao Li, Wen Wang, Xiaochuan Deng, Yun Bai, Hong Chen, and Bo Zhang. "Silicon carbide bipolar junction transistor with novel emitter field plate design for high current gain and reliability." Semiconductor Science and Technology 34, no. 4 (2019): 045001.

[18] Pan, Chen-Wei, and Sheng Cho. "Gate-controlled bipolar junction transistor and operation method thereof." U.S. Patent 10,665,690, issued May 26, 2020.

[19] Chan, Kevin K., Pouya Hashemi, Tak H. Ning, and Alexander Reznicek. "Lateral bipolar junction transistor with abrupt junction and compound buried oxide." U.S. Patent Application 16/502,271, filed October 24, 2019.

[20] Sahu, Abhishek, Abhishek Kumar, and Shree Prakash Tiwari. "Performance Investigation of Universal Gates and Ring Oscillator using Doping-free Bipolar Junction Transistor." In 2020 IEEE Silicon Nanoelectronics Workshop (SNW), pp. 125-126. IEEE, 2020.

[21] Srivastava, Pawan Kumar, Yasir Hassan, Yisehak Gebredingle, Jaehyuck Jung, Byunggil Kang, Won Jong Yoo, Budhi Singh, and Changgu Lee. "Multifunctional van der Waals Broken-Gap Heterojunction." Small 15, no. 11 (2019): 1804885.

[22]     Jain, Aakash Kumar, and Mamidala Jagadesh Kumar. "Sub-10 nm Scalability of Junctionless FETs Using a Ground Plane in High-K BOX: A Simulation Study." IEEE Access 8 (2020): 137540-137548.

[23]     Sahay, Shubham, and Mamidala Jagadesh Kumar. Junctionless field-effect transistors: design, modeling, and simulation. John Wiley & Sons, 2019.

[24]     Devi, L. Beloni, Kundan Singh, Jitendra Kumar, and A. Srivastava. "Triple-sided charged plasma symmetric lateral bipolar transistor on SiGe-OI." Semiconductor Science and Technology 34, no. 5 (2019): 055019.

[25]     Bardeen, J., Brattain, W. H.  The transistor a semi-conductor triode. Phys. Rev. 74, 230-231 (1948)

[26]     Brattain, W. H., Bardeen, J. Nature of the forward current in germanium point contacts. Phys. Rev. 74, 231-232 (1948)

[27]     Raissi, F., Nordman, J. E. Josephson fluxonid diode. Appl. Phys. Lett. 65, 1838-1840 (1994)

[28]     Raissi, F. A brief analysis of the field effect diode and breakdown transistor. IEEE Trans. Electron Dev. 43, 362-365 (1996)

[29]     Sheikhian, I., Raissi, F. Simulation results for nanoscale field effect diode. IEEE Trans. Electron Dev. 54. 613-617 (2007)

[30]     N. Manavizadeh, Raissi, F., Soleimani, A. S., Porfath, M., Selberherr, S. Performance assessment of nanoscale field-effect diodes. IEEE Trans. Electron Dev. 58, 2378-2384 (2011)

[31]     Lee, J. U., Gipp, P. P., Heller, C. M.  Carbon nanotube p-n junction diodes.  Appl. Phys. Lett. 85, 145-147 (2004)

[32]     Amirmazlaghani, M., Raissi, F. Memory cell using modified field effect diode.  IEICE Electronics Exp. 6. 1582-1586 (2009)

[33]     Jafari Toushaee, B., Manavizadeh, N. An Inverter Gate Design Based on Nanoscale S-FED as a Function of Reservoir Thickness. IEEE Trans. Electron Dev. 62, 3147-3152 (2015)

[34]     Sotoudeh, Abolfazl, and Mina Amirmazlaghani. "Graphene-based field effect diode." Superlattices and Microstructures 120 (2018): 828-836.

[35]     Sheikhian, I., and F. Raissi. "High-speed digital family using field effect diode." Electronics Letters 39, no. 4 (2003): 345-347.

[36]     Badwan, Ahmad Z., Zakariae Chbili, Yang Yang, Akram A. Salman, Qiliang Li, and Dimitris E. Ioannou. "SOI field-effect diode DRAM cell: Design and operation." IEEE electron device letters 34, no. 8 (2013): 1002-1004.

[37]     Yang, Yang, Aveek Gangopadhyay, Qiliang Li, and Dimitris E. Ioannou. "Scaling of the SOI field effect diode (FED) for memory application." In 2009 International Semiconductor Device Research Symposium, pp. 1-2. IEEE, 2009.

[38]    Yang, Yang, Akram A. Salman, Dimitris E. Ioannou, and Stephen G. Beebe. "Design and optimization of the SOI field effect diode (FED) for ESD protection." Solid-state electronics 52, no. 10 (2008): 1482-1485.

[39]    Sheikhian, Iraj, and Farshid Raissi. "An improved differential comparator with field effect diode output stage." Journal of Circuits, Systems, and Computers 14, no. 05 (2005): 931-937.

[40]    Lim, W. L., E. J. Moon, J. W. Freeland, D. J. Meyers, M. Kareev, Jak Chakhalian, and S. Urazhdin. "Field-effect diode based on electron-induced Mott transition in NdNiO3." Applied Physics Letters 101, no. 14 (2012): 143111.

[41]    Azizollah-Ganji, Bahram, and Morteza Gholipour. "Effects of the channel length on the nanoscale field effect diode performance." Journal of Optoelectronical Nanostructures 3, no. 2 (2018): 29-40.

[42]    Sze, S. M., Ng, K. K. Physics of Semiconductor Devices. New Jersey: John wiley & sons (2006)

[43]    Hu, Chenming, Modern Semiconductor Devices for Integrated Circuits, New Jersey: Prentice Hall (2010)

[44]    Leblebici, Yusuf, and Sung-Mo Kang. CMOS digital integrated circuits: analysis and design. McGraw-Hill, 1996.

[45]    Rabaey, Jan M., Anantha P. Chandrakasan, and Borivoje Nikolić. Digital integrated circuits: a design perspective. Vol. 7. Upper Saddle River, NJ: Pearson Education, 2003.

(a)

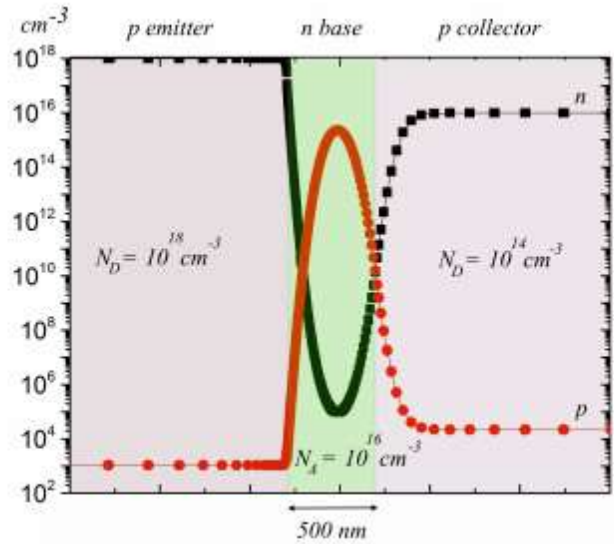

(b)

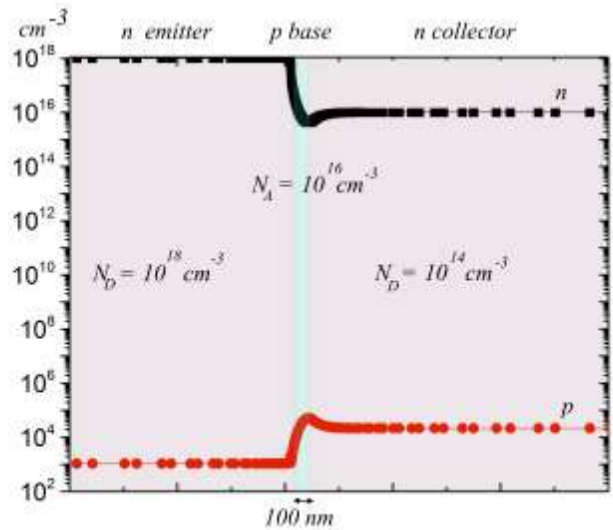

Fig. 1. a) Hole and electron densities for an npn structure made of three regions with donor impurity doping of $10^{18}/cm^3$ and $10^{14}/cm^3$ for emitter and collector regions and acceptor doping density of $10^{16}/cm^3$ for its base. The base width is 500nm. A clear npn structure is created at these dimensions and doping levels. b) The same conditions as that of previous case but now base width is 100nm and it is clear that npn structure is not obtained.

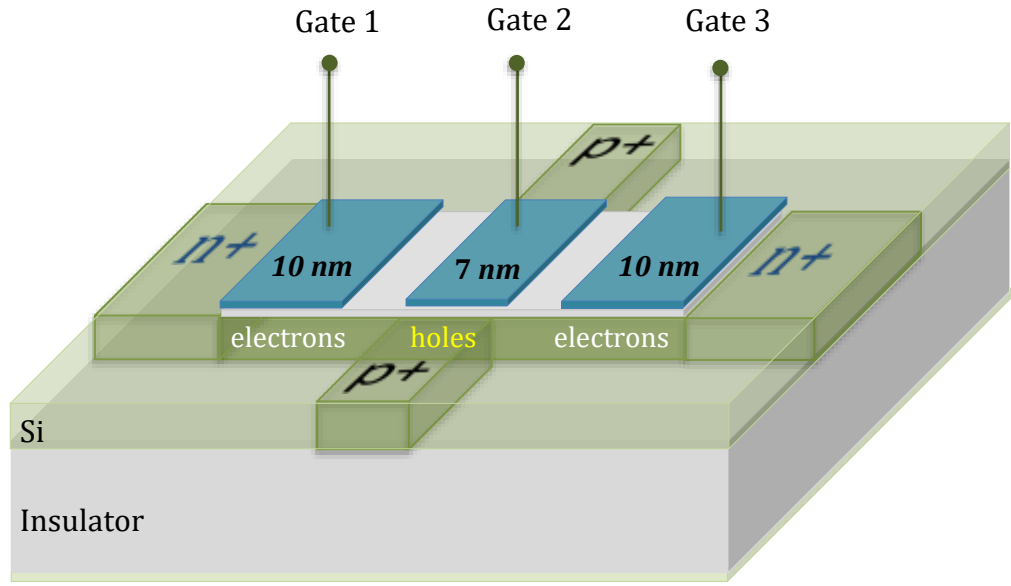

Fig. 2. Schematic of the proposed transistor, which creates the npn regions inside the channel via electric fields of three gates in a silicon-on-insulator substrate at nanometer dimensions. The two side-gates are each 10nm long and the center gate is 7nm. Gate separation of 7nm was used in simulations. The three regions in the channel represent the emitter, base and collector regions of the FEBJT. Contacts to these regions are created by n+ and p+ doped portions along the widths of the source and drain gates and the lengths of the base gate.

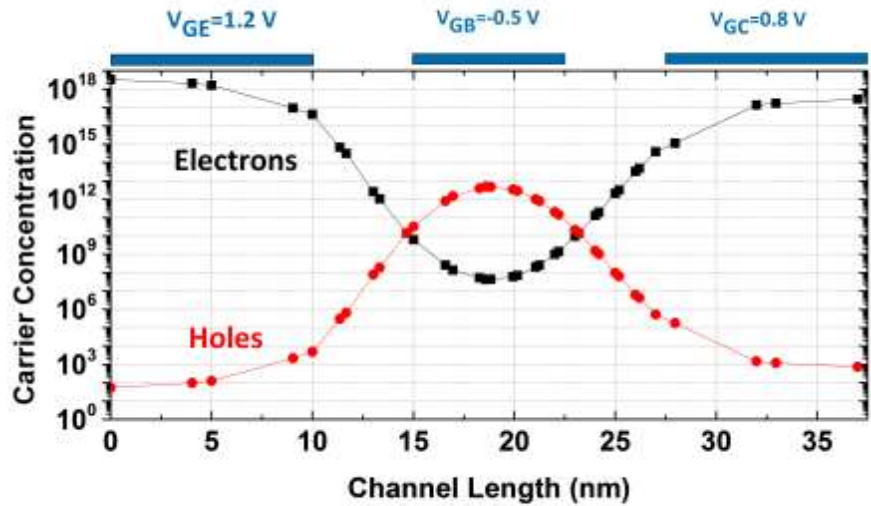

Fig. 3. Carrier densities inside the channel while the total length of the structure from emitter to collector contact is 37nm and the center base gate is only 7 nm. As is observed there exists a clear base region in which holes dominate and are the majority carriers. This is to be compared with Fig. 2.b where the width of the base is 100nm and no npn structure is obtained.

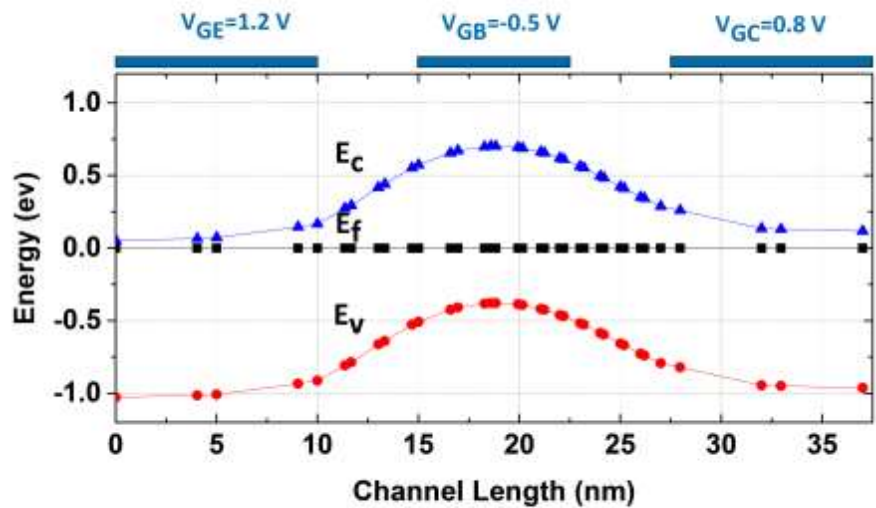

Fig. 4. The band diagram for FEBJT resembling regular BJT band diagram when the base gate width is only 7 nm. This clearly shows the two pn junctions at emitter and collector contacts with the base, which are necessary for transistor behavior. The gate voltages are shown at top and are the same as Fig. 3.

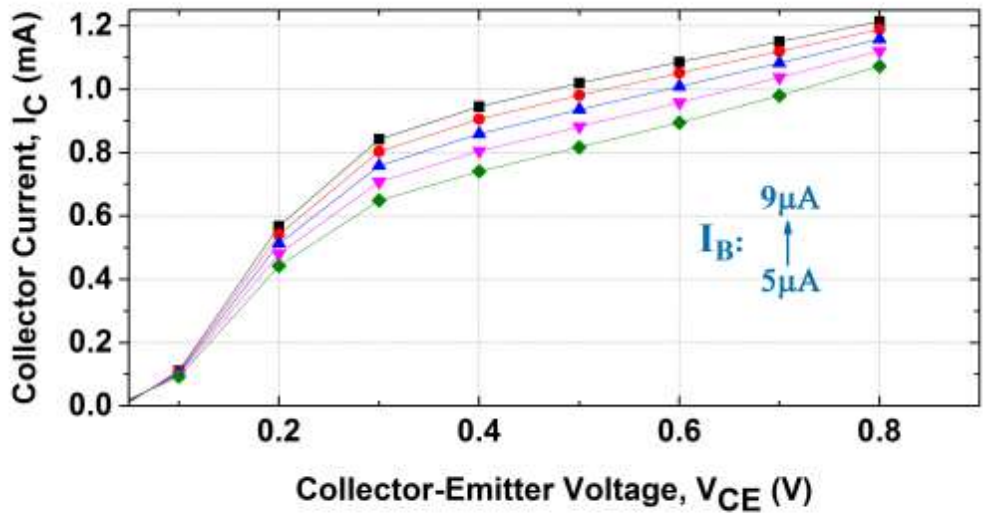

Fig. 5. IV curves of the FEBJT in its forward active mode. The emitter-base diode is forward biased and the base-collector diode is reverse biased. Base current controls the collector current by a gain of about 100 for the particular gate voltages given in previous figures.

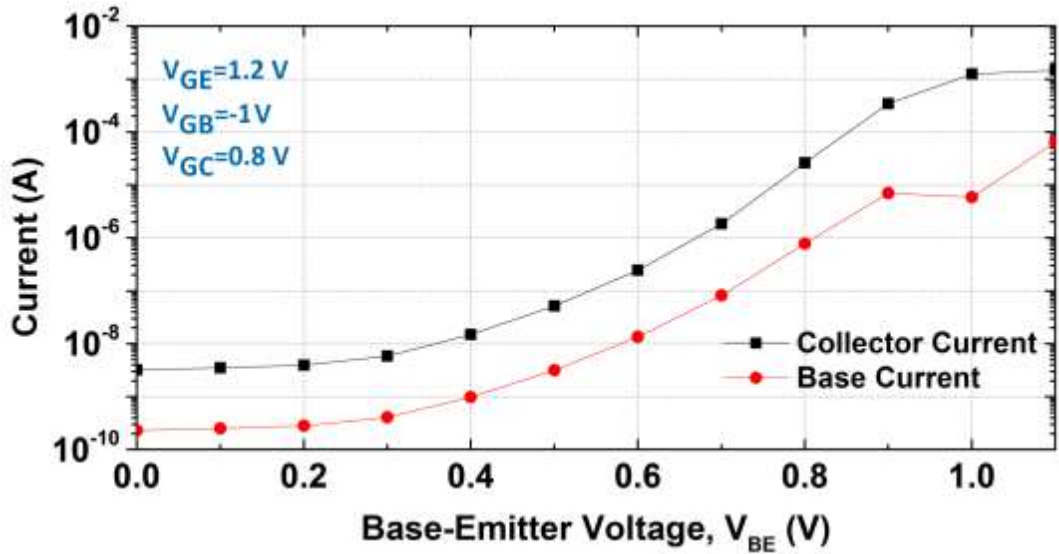

Fig.6. Gummel plot for FEBJT where collector and base currents are shown in semi-log plots versus base-emitter voltage. This plot clearly shows the control base current has on the collector current. Furthermore the slope of the curves at low and high voltages as well as at the center are good indications of physical processes governing the behavior of the device. The three exhibited slopes match what is observed in regular BJTs.

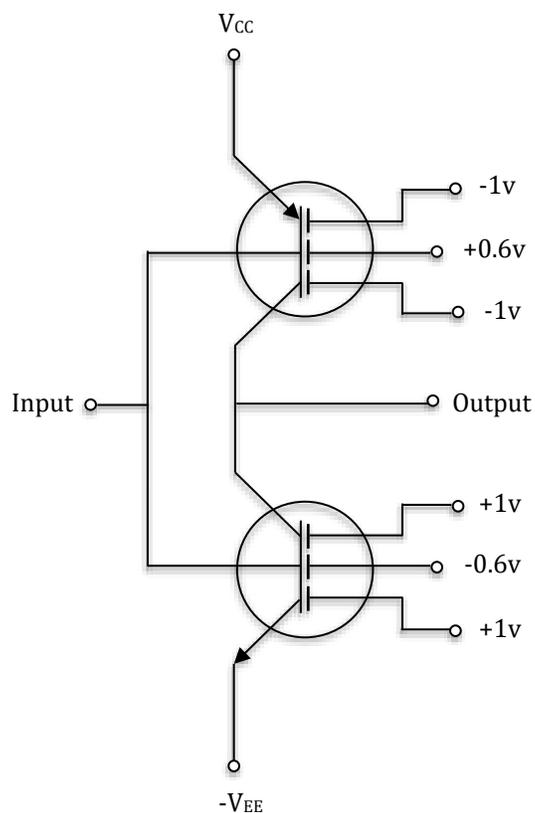

Fig. 7. The inverter NOT gate of FEBJT with base width of 7nm used in simulations. Connections are similar to regular BJT common emitter inverter. Gate voltages are such that a pnp is obtained as the current source connected to maximum positive voltage and a npn for the current sink connected to the minimum negative voltage.

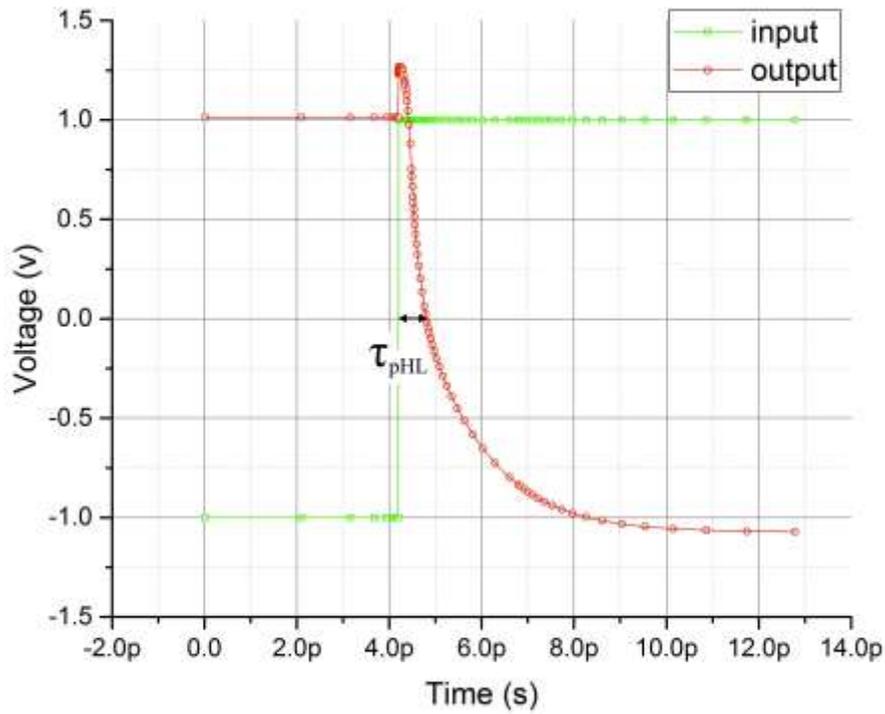

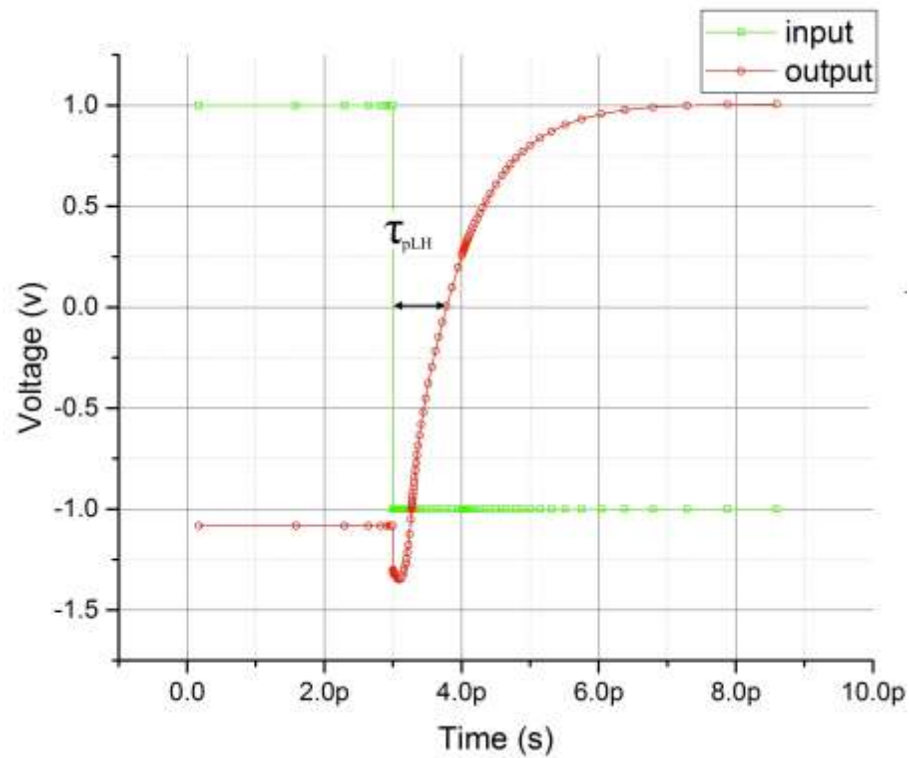

Fig. 8. A square wave is applied to the input of the FEBJT NOT gate and the output is obtained. Rise and fall times indicate a frequency response of 750 GHZ for this inverter amplifier. Extremely fast response is obtained because turning on and off is dominated by diffusion of carriers inside the nano-size base region.

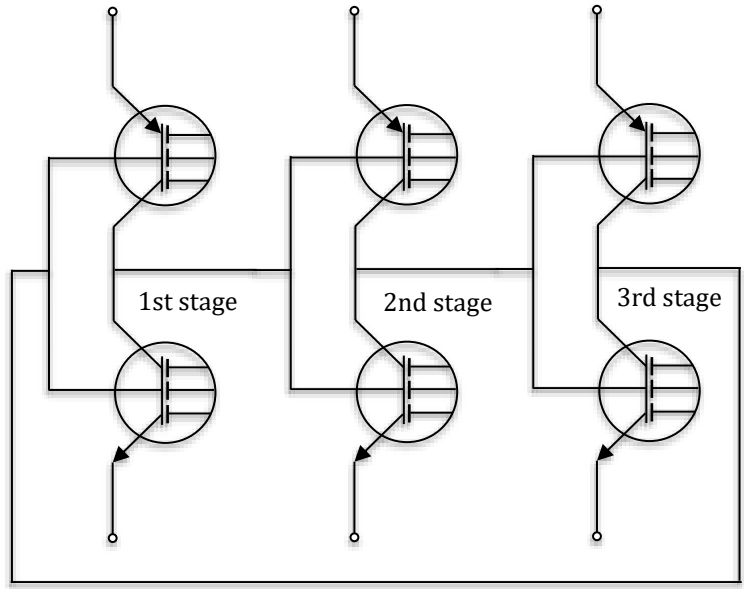

Fig. 9. Three stage ring oscillator using FEBJT. Output of each stage acts as the input of the next stage.

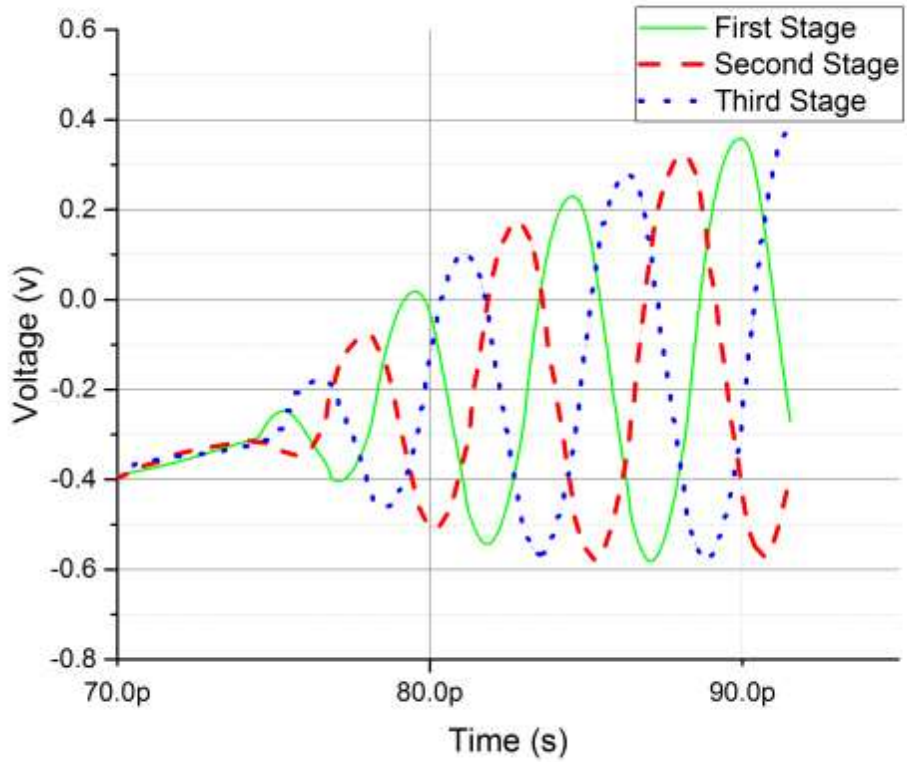

Fig. 10. Oscillation of the three-stage ring oscillator using 7nm base FEBJT. Frequency of oscillation is 245 GHz. In order to start oscillation a ramp voltage was applied to the gates.

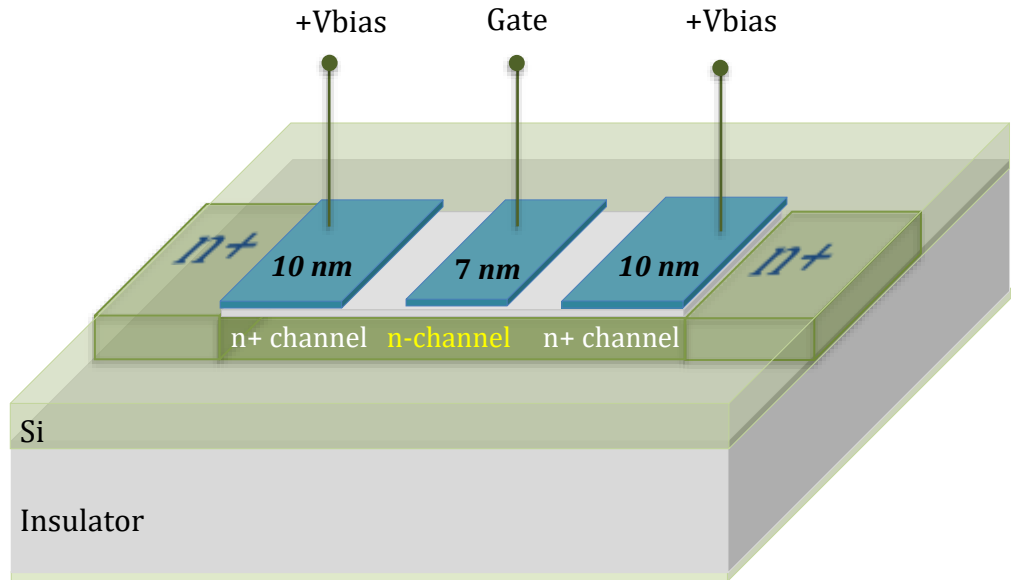

Fig. 11. Gate biasing of the transistor when used as a regular MOS transistor. The voltage applied to the side-gates are kept constant. Center gate plays the role of the input MOS gate. This gate does not overlap with the n+ source and drain. Extra Ion implantation steps to remedy short-channel effects are not necessary.

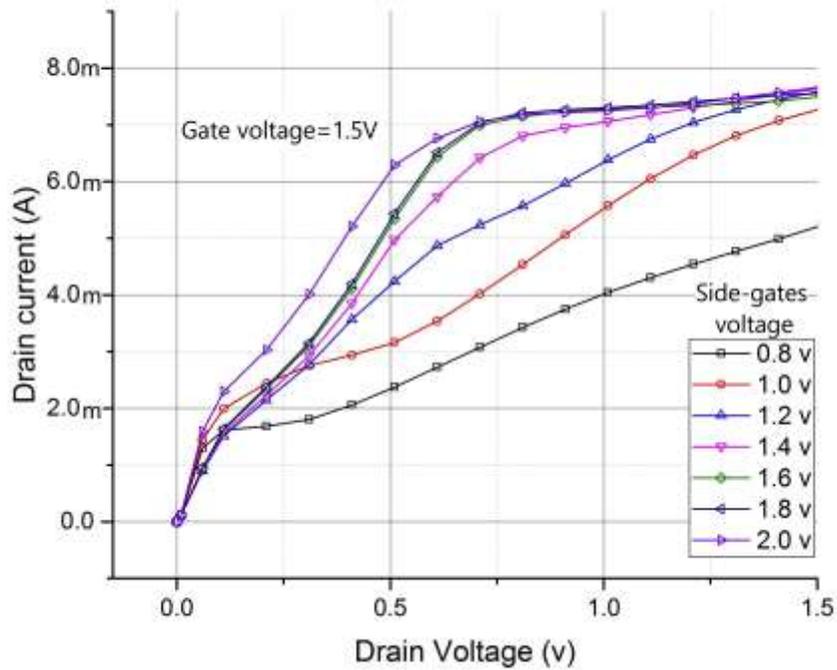

Fig. 12. IV curves when the device is biased as a regular MOSFET. At low side-gate voltages the drain voltage mostly controls the current and there is no saturation regime. As these side-gate voltages increase current assumes more generic MOSFET IV with its saturation region. Better results are obtained if longer side-gates are used.

Table 1. Dimensions and parameters used in simulations.

| Silicon film thickness | 100nm | Si film doping (p) cm$^{-3}$ | $7 \times 10^{17}$ |
|---|---|---|---|
| Buried oxide thickness | 400nm | Emitter contact length | 25 nm |
| Gate oxide thickness | 5 nm | Collector contact length | 25 nm |
| Gates' width | 1µm | Emitter gate length | 10 nm |
| Emitter contact (n) cm$^{-3}$ | $1 \times 10^{19}$ | Collector gate length | 10 nm |
| Collector contact (n) cm$^{-3}$ | $1 \times 10^{18}$ | Base gate length (base width) | 7 nm |

Table 2. Current amplification factor for various base gate voltages.

| $V_{GB}$ (volts) | -1.2 | -1 | -0.9 | -0.8 | -0.7 | -0.6 | -0.5 |
|---|---|---|---|---|---|---|---|
| β | 0.45 | 37 | 75 | 131 | 390 | 2770 | 78000 |